\documentclass[prb,aps,floats,twocolumn]{revtex4}
\usepackage[english]{babel}
\usepackage{amsmath}
\usepackage{amsfonts}
\usepackage{graphicx}
\usepackage{color}
\usepackage{lipsum}

\begin{document}
\title {Josephson effect in topological superconducting rings coupled to a microwave cavity}
\author{Olesia Dmytruk, Mircea Trif and Pascal Simon}
\address{Laboratoire de Physique des Solides, CNRS UMR-8502, Universit\'{e} Paris Sud, 91405 Orsay cedex, France}
\date{\today}

\begin{abstract}
We theoretically study a one dimensional $p$-wave superconducting mesoscopic ring interrupted by a weak link and coupled inductively to a microwave cavity. We establish an input-output description for the cavity field in the presence of the ring, and identify the electronic contributions to the cavity response and their dependence on various parameters, such as the magnetic flux, chemical potential, and cavity frequency. We show that the cavity response is $4\pi$ periodic as a function of the magnetic flux in the topological region, stemming from the so called fractional Josephson current carried by the Majorana fermions, while it is $2\pi$ periodic in the non-topological phase, consistent with the normal Josephson effect. We find  a strong dependence of the signal on the cavity frequency, as well as on the parity of the ground state. Our model takes into account fully the interplay between the low-energy Majorana modes and the gaped bulks states, which we show is crucial for visualizing the evolution of the Josephson effect during the transition from the topological to the trivial phase. 
\end{abstract}

\maketitle

\section{Introduction}

Majorana fermions are zero energy quasiparticles that are their own antiparticles. In condensed matter physics, they emerge as zero energy excitations in the so called topological superconductors, such as genuine $p$-wave superconductors \cite{read2000paired,ivanov2001non,kitaev2001unpaired,yakovenkoEPJB2004}, topological insulators in the presence of conventional superconductors \cite{fu2008superconducting,fu2009josephson,hart2014induced}, nanowire in proximity to $s$-wave superconductor and subject to a magnetic field \cite{oreg2010helical,lutchyn2010majorana,mourik2012signatures,albrecht2016exponential}, in chains of magnetic atoms \cite{nadj2013proposal,klinovaja2013topological, braunecker2013interplay, vazifeh2013self,pientka2013topological,kim2014helical,nadj2014observation}. Majorana fermions in condensed matter physics were first predicted by Kitaev in Ref.~\onlinecite{kitaev2001unpaired} (see Ref.~\onlinecite{alicea2012new} for a review). They obey non-abelian statistics \cite{ivanov2001non} in two dimensions, and they are robust against local perturbations that do not close the superconducting gap, which make them promising candidates as qubits for a functional topological quantum computer.

There are several physical manifestations associated with the presence of Majorana fermions, such as the zero bias anomaly in the topological region, the dependence of the Majorana energy splitting on the chemical potential, Zeeman splitting, or the size of the system, as well as the fractional Josephson effect and the non-abelian statistics pertaining to braiding of the Majoranas~\cite{alicea2012new}. The fractional Josephson effect is one of the hallmarks of the Majorana fermions \cite{kitaev2001unpaired,yakovenkoEPJB2004,fu2009josephson,lutchyn2010majorana,pientka2013signatures}. It corresponds to $4\pi$ periodicity of the supercurrent in the superconductor phase difference across the weak link, as opposed to $2\pi$ periodicity for a usual Josephson effect in conventional $s$-wave superconductors, and it arises because of the degeneracy of the zero energy Majorana modes. In other words, because of such degeneracy, there is a coherent charge $e$ transfer across the weak link, instead of $2e$ as in the normal Josephson effect. There are various ways to reveal the fractional Josephson effect, such as the measurement of the Shapiro steps in the $I-V$ characteristics of a voltage biased junction \cite{rokhinson2012fractional,wiedenmann2016}, by employing a  resistive shunted SQUID setup~\cite{della2007measurement}, or by performing microwave spectroscopy of the junction. The latter method is particularly interesting as it allows to access also the exited states of the junction. There are several theoretical works that study the interplay of the Majorana fermions physics and microwaves in a superconducting cavity QED setup
~\cite{trif2012resonantly,schmidt2013majorana,schmidt2013njp,cottet2013squeezing,muller2013detection,XuePRA2013, ohm2014majorana,XueSciRep2015,yavilberg2015fermion,ohm2015microwave,dmytruk2015cavity}. In contrast to the usual electronic methods, this approach is unique in that it can be totally non-invasive, i.e. it does not alter the electronic system, a crucial feature in view of using these excitations for quantum computing. However, almost all of these studies deal with effective models that consider {\it only} the Majorana fermions interacting with the cavity field, which cannot account for the evolution of the Majorana fermions through the topological transition from a non-trivial to a trivial topological superconductor. More specifically, the Majorana fermions emerge as edge modes in a topological superconductor: below the phase transition the system is non-trivial, and shows edge modes, while above the phase transition the system is trivial, and is has no edge modes. Majorana fermions are thus intimately related to the bulk physics, and are not simply impurity states in the superconductor. The fractional Josephson effect disappears above the phase transition, where it becomes the normal $2\pi$ periodic one. In order to account for the evolution of the system from topological to trivial, associated with the fractional and normal Josephson effect, respectively, one needs to consider the Majorana and the bulk states on the same footing. In this paper, we do precisely that in the context of the cavity QED setup that was first proposed in Ref~. \onlinecite{blais2004cavity} and experimentally implemented in Ref.~\onlinecite{wallraff2004strong}. The recent progress in engineering sensitive superconducting cavities makes it possible to implement on-chip mesoscopic circuits
coupled to high finesse resonators~\cite{samkharadze2015high}.

\begin{figure}[t] 
\centering
\includegraphics[width=0.9\linewidth]{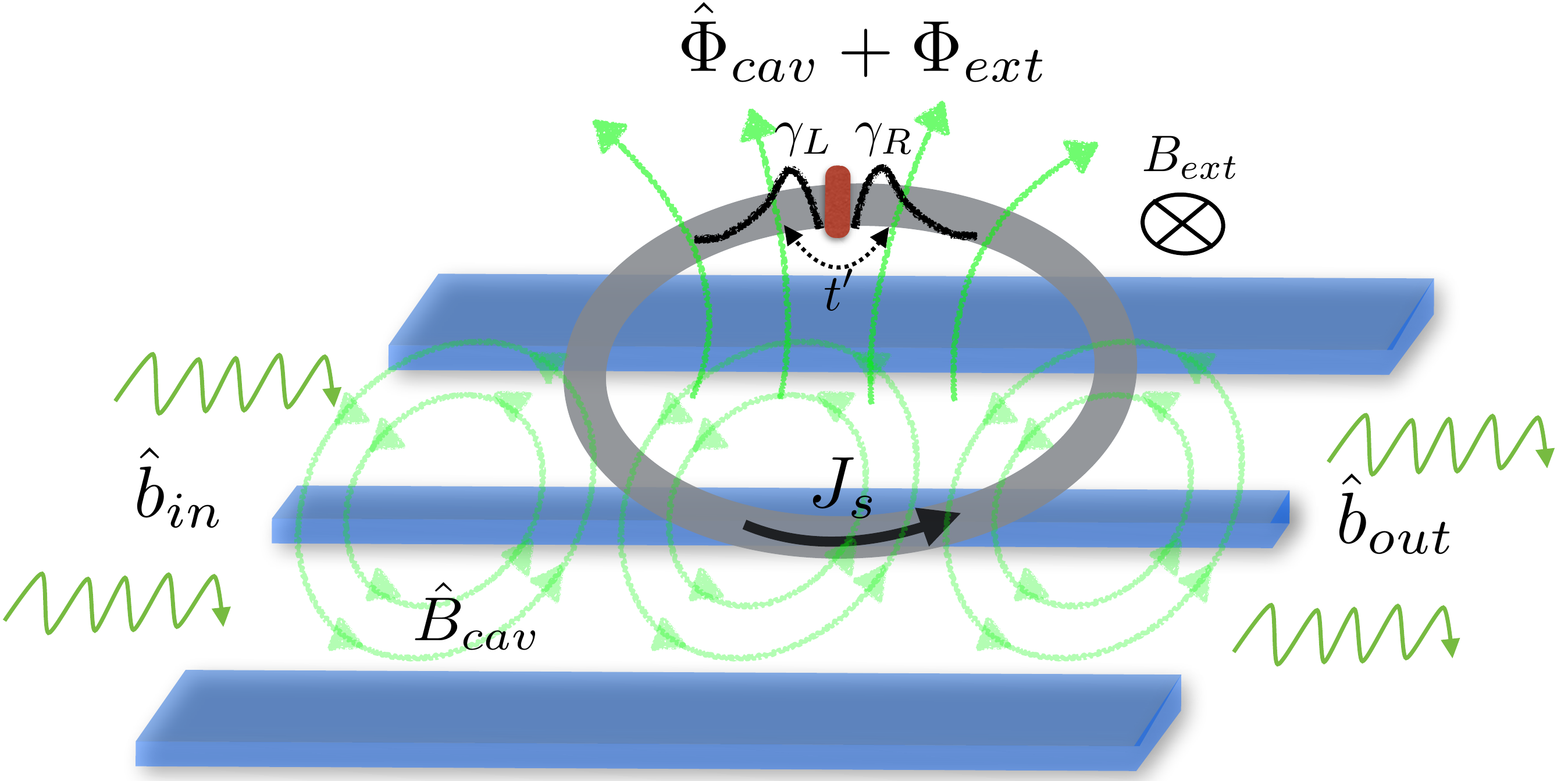}\vspace{0.5cm}
\includegraphics[width=0.95\linewidth]{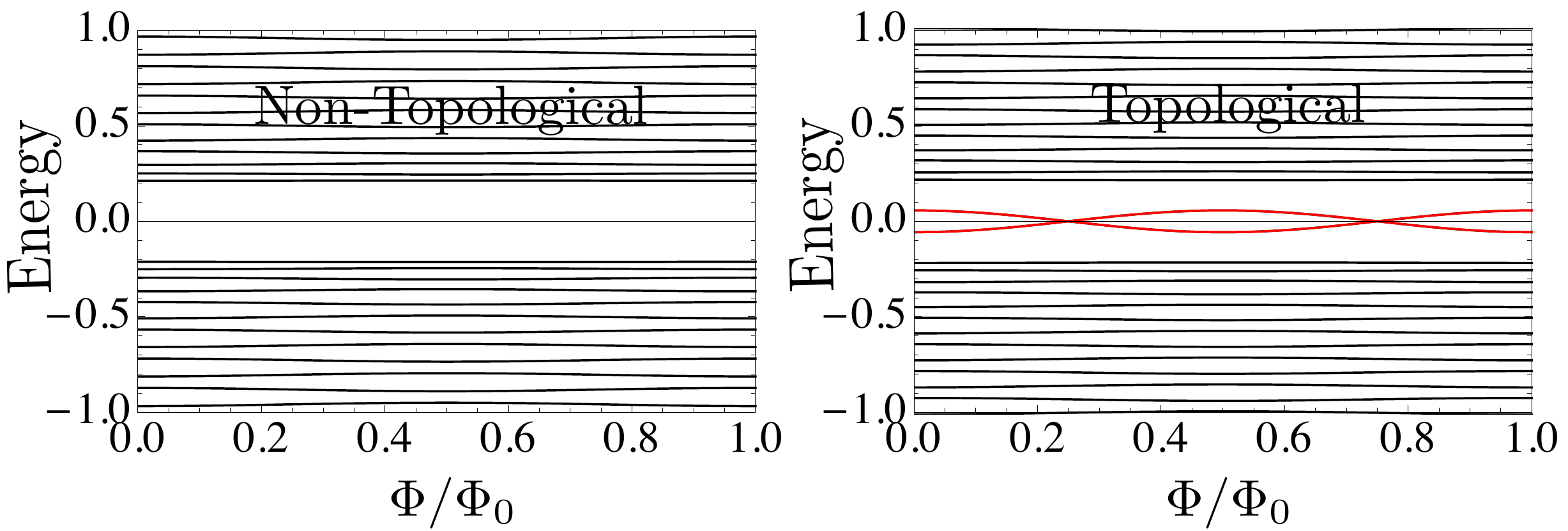}
\caption{Top: A sketch of the combined topological superconducting ring inserted in a microwave cavity. The magnetic field of the cavity $\hat{B}_{cav}$ pierces the ring and gives rise to a fluctuating magnetic flux. A perpendicular, external magnetic field $B_{ext}$ is also applied, so that the total flux felt by the electrons in the ring is $\Phi_{tot}=\Phi_{ext}+\hat{\Phi}_{cav}$. The ring is interrupted by a weak link, that pertains to the hopping strength $t'\ll t$ ($t$ is the hopping strength within the ring, see the text). In the topological region, there are Majorana fermions on the left ($\gamma_L$) and on the right ($\gamma_R$), and a persistent current $J_s$ is flowing within the ring. The cavity is probed in the input-output method, and the output signal $\hat{b}_{out}$ is to be compared with the input one $\hat{b}_{in}$. Bottom: Spectrum of the ring as a function of the flux $\Phi_{ext}$ in the non-topological (left) and topological (right) regions, respectively. In red are depicted the Majorana energy levels that show oscillations with a period $\Phi_0/2$.} 
\label{fig:scheme} 
\end{figure}

In this paper, we study a mesoscopic superconducting ring with a weak link threaded by a magnetic flux and coupled to the cavity. We consider an inductive coupling between the mesoscopic ring and the cavity~\cite{mendes2015cavity}. We show that the electronic susceptibility has a different periodicity as a function of the magnetic flux in the topological and non-topological regions, thus allowing us to probe whether the wire is in topological or non-topological region: $2\pi$ ($\Phi_0/2=h/2e$) periodicity of the Josephson current~\cite{pientka2013signatures} corresponds to the non-topological region, while $4\pi$ ($\Phi_0=h/e$) periodicity of the observables corresponds to the topological region, when Majorana fermions emerge in the system.

The paper is organized as follows: in Sec.~II we introduce the $p$-wave superconducting model and the Hamiltonians. Specifically, we consider the Kitaev model in the topology of a ring with a weak link. Moreover, we assume the ring to be threaded by both $dc$ and $ac$ magnetic fluxes. In Sec.~III we analyze the coupling between the cavity and the superconducting ring, as well as calculate the response of the cavity due to this coupling. In Sec.~IV we calculate the reactive and dissipative response of the cavity, and show the numerical results. Finally, in Sec.~V we end up with conclusions and provide some outlook.    

\section{The system and model Hamiltonian} 

In Fig.~\ref{fig:scheme}, we present a sketch of the system under consideration: a topological superconducting ring which contains a weak link at some position coupled to a superconducting microwave cavity. The ring is subject to both a dc magnetic flux, and an ac (quantum) flux from the cavity, which gives rise to a coupling between the two systems, the ring (electrons) and cavity (photons). In this paper, we consider the Kitaev model Hamiltonian as  describing the electronic $p$-wave superconductor. Such a model could be implemented experimentally utilizing  a semiconducting nanowire with spin-orbit coupling in the proximity of $s$-wave superconductor and subject to a magnetic field~\cite{oreg2010helical,lutchyn2010majorana}. We assume the ring is interrupted at some point by a weak link that emulates the  Josephson effect in a SIS junction. The external magnetic flux allows for a continuous phase change across the weak link, which controls the Josephson current. On the other hand, the fluctuating cavity flux couples to the flux-biased system and monitors its dynamics. 

The general form of the combined Hamiltonian reads:
\begin{align}
H_{sys}&= H_{el} + H_{ph}+H_T\,,\\
\hspace{-1cm}H_{el}&=-\sum_{j = 1}^{N - 1}(te^{i\Phi_{jj+1}} c_j^\dag c_{j+1}+ \Delta e^{i\phi_{jj+1}} c_j c_{j+1} + {\rm h. c.})\nonumber\\
&-\mu\sum\limits_{j = 1}^{N}c^\dag_jc_j\,,\\
H_T&=-t'e^{i\Phi_{N1}}c_N^\dagger c_1+ {\rm h. c.}\,,
\end{align}
where $H_{el}$ is the Hamiltonian of the $p$-wave superconductor, $H_{T}$ is the tunneling Hamiltonian between the ends in the presence of the magnetic flux,  and  $H_{ph} = \omega_c a^\dag a$ is the Hamiltonian of a single mode cavity, where  $\omega_c$ is the cavity frequency and $a^\dag (a)$ is the photon creation (annihilation) operators, respectively. Also, $t$ is the hopping parameter, $\Delta$ is the $p$-wave superconducting  pairing potential, $\mu$ is the chemical potential, and $N$ is the total number of sites.  $c_j^\dag (c_j)$ is the electron creation (annihilation) operators, respectively, at the site $j$. Moreover, $\Phi_{jj+1}$ and $\phi_{jj+1}$ are the phases on the tunneling matrix elements $t$ and the superconducting pairing $\Delta$ at positions $j$ caused by the external fluxes. They read:
\begin{align}
\Phi_{jj+1}&=\frac{e}{\hbar}\int_{j}^{j+1}dx A(x)\equiv\frac{2\pi\Phi_{tot}}{N\Phi_0}\,,\\
\phi_{jj+1}&=\frac{4\pi\Phi_{tot}j}{N\Phi_0}\,.
\end{align} 
The form of the second term (the phase on the pairing $\Delta$) can be found from utilizing  the following assumption: the $p$-wave pairing in the Kitaev chain is assumed to be induced by an $s$-wave superconductor underneath the wire that is interrupted at the link. In such a case, we can consider that the instantaneous supercurrent flowing through the superconductor vanishes, namely:
\begin{equation}
J_s=\frac{2e}{m}|\psi|^2(\hbar\nabla\phi-2eA)\equiv0\,,
\end{equation}
where $m$, $|\psi|^2$, and $\phi$ are the electronic mass, the density of superconducting electrons in the $s$-wave superconductor, and  its phase, respectively. The latter is directly imprinted into the wire by means of proximity effect.  We can easily solve this differential equation, which then gives the phase on the $p$-wave superconductor~\cite{pientka2013signatures} $\phi(j)\equiv\phi_{jj+1}=4\pi\Phi_{tot} j/N\Phi_0$. 

We note that the system is in topological (non-topological) phase when $\left|\mu\right|<2t$ ($\left|\mu\right|>2t$), and it supports Majorana zero energy end modes (no Majorana end modes). 

The magnetic flux contains both the dc flux $\Phi_{dc}$, as well as a time-dependent component $\hat{\Phi}_{ac}(t)$, or $\Phi=\Phi_{dc}+\hat{\Phi}_{ac}(t)$. We assume that the ac flux is due to the cavity photons, and it is given by $ \hat{\Phi}_{ac}=i\lambda(a-a^\dag)$, with $\lambda$ being the coupling constant between the ring and the cavity. For the moment, we keep this parameter arbitrary, and discuss its possible values later on.  

We can rewrite the full Hamiltonian in a simpler form by performing a unitary transformation $U=\exp{[-2\pi i\sum_{j=1}^N(j-1/2)n_j\Phi_{dc}/N\Phi_0]}$ so that the entire dependence on the dc flux $\Phi_{dc}$ is transferred to the weak link tunneling Hamiltonian $H_T$, or:
\begin{equation}
H_{T}\rightarrow U^\dagger H_{T}U=-t'e^{i2\pi\Phi_{dc}/\Phi_0}e^{i2\pi \hat{\Phi}_{cav}/N\Phi_0}c_N^\dagger c_1+ {\rm h. c.}
\end{equation}
We mention that such a transformation can only gauge away the dc flux (i.e. transfer it entirely to $t'$), while the ac (or cavity component) cannot be gauged away in a simple way because such a transformation would not commute with the photonic Hamiltonian $H_{ph}$. However, we assume that $\hat{\Phi}_{ac}\ll\Phi_0$, so that we can expand the exponentials in $\hat{\Phi}_{ac}$ up to second order in this quantity.    
With these approximations, the final Hamiltonian can be written as the sum of the electrons (in the presence of the dc flux), the photons, and their mutual interaction:
\begin{align}
H_{sys}&\approx H_{el}+H_T+H_{ph}-\hat{\Phi}_{ac}I+\frac{1}{2}(\hat{\Phi}_{ac})^2D\,,\\
I&=\frac{2\pi i}{N\Phi_0}\sum_{j=1}^{N-1}\Big[tc^\dagger_jc_{j+1}+2j\Delta c_j c_{j+1}\nonumber\\
&+t'e^{i2\pi\Phi_{dc}/\Phi_0}c^\dagger_Nc_1-{\rm h. c.}\Big]\,,\\
D&=-\left(\frac{2\pi i}{N\Phi_0}\right)^2\sum_{j=1}^{N-1}\Big[tc^\dagger_jc_{j+1}+(2j)^2\Delta c_j c_{j+1}\nonumber\\
&+t'e^{i2\pi\Phi_{dc}/\Phi_0}c^\dagger_Nc_1+{\rm h. c.}\Big]\,.
\end{align}
This is the Hamiltonian that we will be utilizing in the following to describe the response of the cavity, up to second order in the electron-cavity coupling.   

Such configuration (ring geometry) have been studied in the past in the context of persistent currents in {\it normal} rings, and it was shown to be physically equivalent to a SNS junction, where the low-energy system is described by the so called Andreev levels. In both cases, the spectrum is quantized due to the finite size of the ring (N system), or the finite size of the N part (SNS junction). We mention that for the SNS junction, if the length of the normal part $L\ll\xi$ ($\xi$ is the coherence length), the spectrum shows a gap that is modulated by the external flux, and can even vanish. This gap, or mini gap, is usually much smaller than the bulk superconducting gap, and these levels are responsible for all the low-energy transport properties of the junction. The Andreev levels in the presence of the flux give rise to persistent currents which, for the state $n$ with  energy $\epsilon_n$ reads:
\begin{equation}
i_n=-\frac{\partial\epsilon_n}{\partial\Phi_{dc}}\,.
\end{equation}        
The Josephson current carried by these states is then given by $I_J=\sum_{n}f_ni_n$, where $f_n$ are the occupations (Fermi-Dirac) of the level $n$.

\section{Electronic susceptibility and cavity response}

\subsection{Theoretical approach to susceptibility}

In this section, we discuss the cavity response in the presence of the $p$-wave superconductor, and the dependence on various electronic parameters, such as the chemical potential and the applied magnetic flux $\Phi_{dc}$.  

There are two ways to approach the problem: utilizing a quantum description, namely treating the photons as quantum objects, or assuming the ac component of the flux is a classical oscillating quantity, of the form $\Phi_{ac}=\Phi_{ac}(0)\sin(\omega t)$. In this paper, we will employ the first method, but compare with the expected results utilizing a classical description. The reason for doing so is that the first method is suited to also enter the quantum regime, where the coupling between the photons and the electronic system is strong and one needs to employ a polaritonic-like description. While such a case is left for a future study, we believe that the formalism developed in this paper will be of great usefulness.    

In previous works \cite{dmytruk2015cavity,dmytruk2016out}, we derived explicitly the response of the cavity due to its coupling to an electronic system. Here we give a brief summary of the derivation, and depict the results for the particular type of couplings found here. For more details on the general derivation we refer the reader to Ref.~\onlinecite{dmytruk2015cavity}. The basic idea is the following: the cavity is probed in transmission, namely photons are sent toward it, and the outcome photons are collected. By comparing the input and output signals, one can infer the properties of the electronic system. One can start from the equation of motion for the cavity field~\cite{clerk2010introduction}:
\begin{align}
\dot{a}=\frac{i}{\hbar}[H_{sys},a]-\frac{\kappa}{2}a-\sqrt{\kappa}b_{in}\,,
\label{eq_motion_in}
\end{align}
for the input field, and
\begin{align}
\dot{a}=\frac{i}{\hbar}[H_{sys},a]+\frac{\kappa}{2}a-\sqrt{\kappa}b_{out}\,,
\label{eq_motion_out}
\end{align}
for the output field, with $b_{out}=b_{in}+\sqrt{\kappa}a$. Here,  $\kappa$, $b_{in}$, and $b_{in}$ are the  escape rate of the photons from  the cavity, the input, and the output photonic fields, respectively, while $H_{sys}$ is the total Hamiltonian of the cavity and electrons, including their interaction. This is in general a very complicated problem to solve, as the photonic and electronic dynamics are correlated and thus there are back-action effects between the two systems. However, in the weak coupling limit the problem can be solved iteratively. Note that the second order contribution to the electron-photon coupling can be written as  $(a^\dagger)^2+a^2+2n_{ph}+1$, with $n_{ph}\equiv a^\dagger a$, and we see that the third and fourth terms renormalize the cavity frequency and the weak tunneling amplitude, respectively. The first two terms correspond to creation or annihilation of two photons, and we can disregard them in the weak coupling limit. With this approximation, we can insert this expression into the equation of motion for the photonic operator in Eqs.~(\ref{eq_motion_in}) and  (\ref{eq_motion_out}), switch to the frequency domain, and solve these equations iteratively up to second order in $\lambda$.  By taking the average over the unperturbed (or uncoupled to the cavity) electronic Hamiltonian, we obtain:
\begin{align}
-i\omega a&=-i(\omega_c+\lambda^2\langle D\rangle)a-\frac{\kappa}{2}a-\sqrt{\kappa}b_{in}\nonumber\\
&-i\lambda^2\Pi(\omega_c)a+\lambda\langle I\rangle\,,
\end{align} 
where $\Pi(\omega)=\int dt\exp{(i\omega t)}\Pi(t)$ with 
\begin{align}
\Pi(t)=-i\theta(t)\langle[I(t),I(0)]\rangle\,.
\end{align}
Note that all the expectation values are taken over the electronic system in the absence of the cavity.  Let us discuss briefly the resulting equation of motion and further approximations. We assume the input field $b_{in}$ contains a large number of photons, and thus we can neglect the last term that acts as an extra input source (and which is entirely due to the current flowing through the weak link). The opposite limit, when the input field is only due to vacuum fluctuations, can in principle lead to emission of photons due to current fluctuations in the ring. Such physics have been studied in connection with light emission by voltage biased tunnel and Josephson junctions, and is beyond the scope of our paper.

With this, we see that the coupling to the superconductor affects the cavity in two ways: by changing its frequency, and by changing its quality factor (or the escape rate $\kappa$). More precisely, we have:
\begin{align}
\delta\omega&=\lambda^2\left[\langle D\rangle+{\mathcal Re}\Pi(\omega_c)\right]\,,\\
\kappa'&=\kappa-\lambda^2{\mathcal Im}\Pi(\omega_c)\,.
\end{align}      
In Ref.~\onlinecite{dmytruk2015cavity,dmytruk2016out} we showed explicitly that the transmission $\tau$ of the cavity can be written as follows:
\begin{equation}
\tau\equiv\frac{b_{out}}{b_{in}}=\frac{\kappa}{-i(\omega-\omega_c)+\kappa+i\lambda^2\Pi_{tot}(\omega_c)}\,,
\end{equation}   
where $\Pi_{tot}(\omega_c)=\Pi(\omega_c)+\langle D\rangle$. It is instructive to rewrite the two terms  in more explicit forms:
\begin{align}
\langle D\rangle&=\sum_{n}f_n\langle n|D|n\rangle\,,\\
\Pi(\omega)&=\sum_{n\neq m}|\langle n|I|m\rangle|^2\frac{f_n-f_m}{-(\epsilon_n-\epsilon_m)+\omega+i\eta}\,,
\end{align}
where $f_n$ is the occupation of the electronic  energy state $\epsilon_n$, and $\eta$ is a small number that account for the dissipation. We can rewrite  $\Pi_{tot}(\omega)$ in a more transparent form by utilizing the following sum rule \cite{trivedi1988mesoscopic,ferrier2013phase}:
\begin{equation}
\langle n|D|n\rangle+\sum_{m\neq n}\frac{|\langle n|I|m\rangle|^2}{\epsilon_n-\epsilon_m}=\frac{1}{2}\frac{\partial^2\epsilon_n}{\partial\Phi^2}\,,
\end{equation}
so that we obtain
\begin{align}
\Pi_{tot}(\omega)&=\frac{\partial J_{s}}{\partial\Phi}+\sum_n\frac{\partial f_n}{\partial\Phi}\frac{\partial\epsilon_n}{\partial\Phi}\nonumber\\
&-\omega\sum_{n\neq m}\frac{f_n-f_m}{\epsilon_n-\epsilon_m}\frac{|\langle n|I|m\rangle|^2}{(\epsilon_n-\epsilon_m)-\omega-i\eta}\,,
\end{align} 
where $J_s=\sum_{n}f_ni_n$ is the persistent (or Josephson) current flowing through the ring in the presence of a $dc$ flux. In this form, the expression for the susceptibility that is measured by the cavity (via its transmission) has the same form as the one derived previously for normal \cite{trivedi1988mesoscopic} and for superconducting rings \cite{dassonneville2013dissipation,ferrier2013phase}. However, in those previous works they utilized a classical description of the $ac$ perturbation. Moreover, they assumed the electronic system is not closed, but coupled to an external bath which leads to a finite width of the electronic levels. To make correspondence between their results and ours in such a case, we can write:
\begin{align}
\Pi_{tot}(\omega)&=\frac{\partial J_{s}}{\partial\Phi}+\frac{\omega}{\omega+i\gamma}\sum_n\frac{\partial f_n}{\partial\Phi}\frac{\partial\epsilon_n}{\partial\Phi}\nonumber\\
&-\omega\sum_{n\neq m}\frac{f_n-f_m}{\epsilon_n-\epsilon_m}\frac{|\langle n|I|m\rangle|^2}{(\epsilon_n-\epsilon_m)-\omega-i\gamma}\,,
\end{align} 
where  $\gamma$ is the relaxation rate of the levels (assumed, for simplicity, to be the same for all levels). We see that, in principle, there are three contributions to the cavity response:  the persistent current or non-dissipative, the diagonal or the decay of the levels, and the non-diagonal stemming from the usual Kubo contribution, respectively. In the following, we assume the zero temperature limit ($T=0$), which in turn implies that the second term vanishes. The first term is independent on the frequency $\omega$, and thus any dependence on this parameter will be due to the last term.    

\begin{figure}[t] 
\begin{minipage}[t]{0.99\linewidth}
\includegraphics[width=\linewidth]{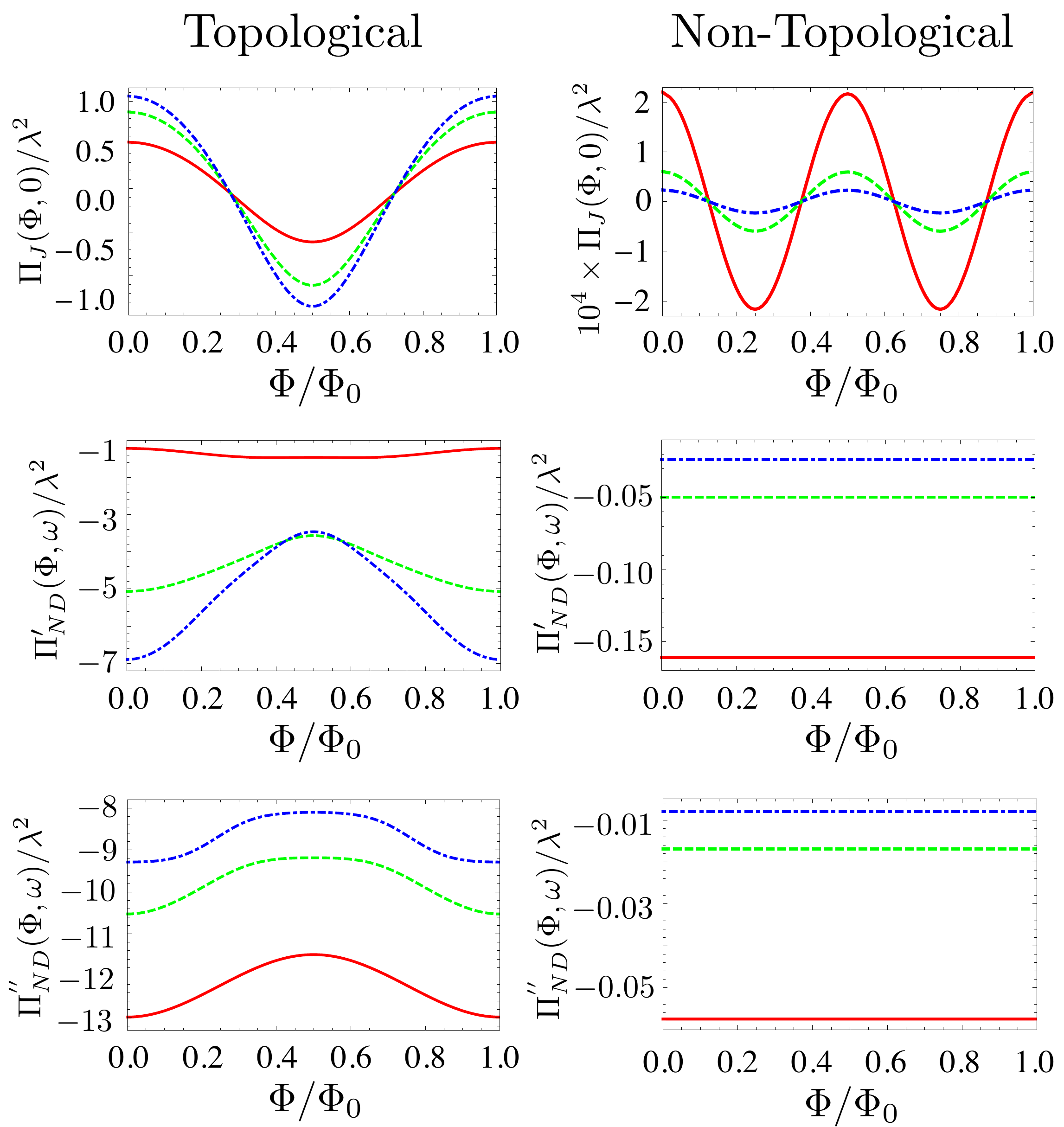}
\end{minipage}
\caption{Left row (right row), from top to bottom: Dependence of the Josephson, non-diagonal real, non-diagonal imaginary susceptibilities respectively,  in the topological (non-topological) region on the magnetic flux $\Phi/\Phi_0$  for $f_M=0$. The red, green, and blue lines  correspond to $\mu=-1.5$ ($\mu=-2.5$),  $\mu=-1$ ($\mu=-3$),  and  $\mu=-0.5$ ($\mu=-3.5$), respectively. The phase transition point is  at $\mu=-2$. The parameters for the plots  are  $\Delta=0.1$, $t'=0.2$, $N = 50$, and all energies are expressed in terms of $t=1$.} 
\label{fig:kitaev_chain_pull}
\end{figure}

For the classical derivation of the response we refer the reader to Refs.~[\onlinecite{trivedi1988mesoscopic,ferrier2013phase,sticlet2014dynamical}]. A few comments are in order. In the case of a SNS junction, the entire contribution to the cavity signal is coming from the sub gap  Andreev states in the normal region, while any bulk effects can be neglected since the bulk gap is much larger than the minigap. That need not be the case for a $p$-wave superconductor, as this system can suffer a topological transition and close its bulk gap, which in turn should affect strongly the cavity response as a function of the magnetic flux. Thus, the energy levels in the expression for the susceptibility are the ones that correspond to the entire bulk spectrum, and not only the mini gap states. 

\subsection{Flux dependence of the cavity response}

In Fig.~\ref{fig:kitaev_chain_pull}, we plot separately the different contributions to the susceptibility measured by the cavity, as a function of the external flux: the Josephson current contribution (top), the real  (middle) and the imaginary (bottom)  non-diagonal contributions both in the topological (left column) and the non-topological phases (right column). We see that in the topological regime, all susceptibilities are $\Phi_0$ periodic, with large amplitudes variations. The amplitude of the $\Phi_0$ oscillations, however, diminishes as the system approaches the topological transition, and disappear after this transition, with only the  $\Phi_0/2$ becoming  manifest. Such $\Phi_0/2$ oscillations pertain to the normal Josephson effect, and are much smaller in amplitude, as can be seen from the top plot on the right column. We note that such oscillations exist also for the non-diagonal terms, but they are too small to be depicted (of the order $~10^{-6}$). 

Let us now discuss the various terms contributing to the total susceptibility, as well as the effect of Majorana fermions on these quantities. As mentioned already, in the topological regime, there are Majorana fermions emerging at the edge of the ring, and which are coupled via the weak link.  In a low-energy description, these Majoranas are responsible for the so called fractional Josephson effect, one of the hallmarks of the Majorana fermions physics. However, the response of the cavity is sensitive not only to the presence of  Majoranas, but also to the interplay between these excitations and the bulk states of the superconductor, in particular close to the topological phase transition. The first term is insensitive to such transitions, as it is given by the derivative of the persistent current with respect to the applied flux. The second term instead contains the matrix elements between the Majorana  and the bulk states, it has both a real and imaginary parts, and depends on the cavity frequency $\omega_c$. We can thus decompose it as follows: 
\begin{align}
\Pi_{tot}(\omega)=\Pi_{BB}(\omega)+\Pi_{BM}(\omega)+\Pi_{MM}(\omega),
\end{align}
where $\Pi_{BB}$ is the part of the susceptibility that comes from the transitions between (different) bulk levels, $\Pi_{BM}$ corresponds to the transitions between bulk and Majorana levels while $\Pi_{MM}$ corresponds to the transitions between Majorana levels and $\Pi_{MM}(\omega)\equiv0$ in the present setup. Note that in the non-topological regime $\Pi_{BM}=0$, as there are no Majorana states present.

\begin{figure*}[t] 
\includegraphics[width=0.99\textwidth]{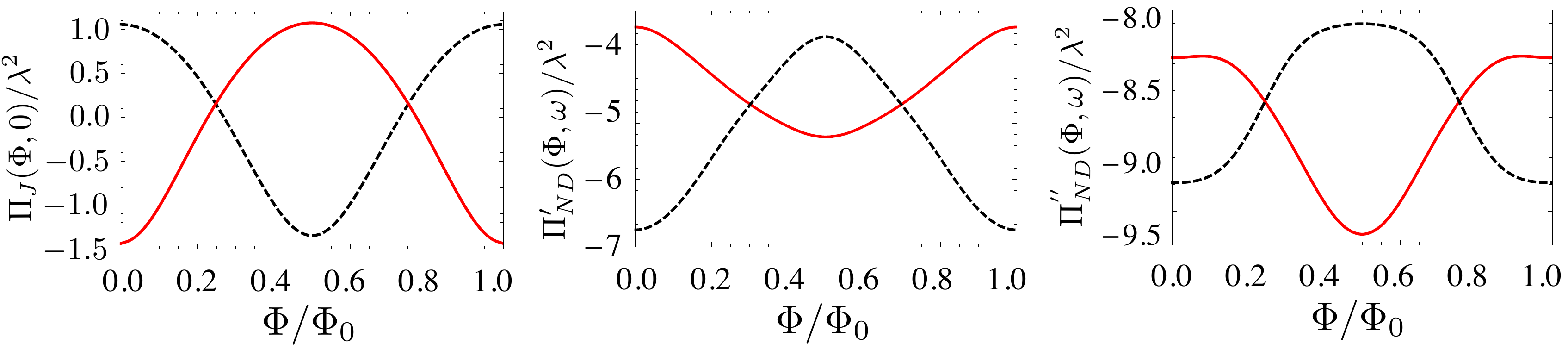}
\caption{Dependence of the Josephson, non-diagonal real, and non-diagonal imaginary susceptibilities, respectively, on the magnetic flux $\Phi$ and parity $f_M$. The dashed-black (full-red) curves correspond to parity $f_M=0$ ($f_M=1$). The oscillations of the susceptibilities for  the two parties  are both shifted by $\Phi_0$, as well as asymmetric in amplitudes (once superimposed).  The parameters for the plots  are $\mu=-0.5$,  $\Delta=0.1$, $t'=0.2$, $N = 50$, and all energies are expressed in terms of $t=1$.} 
\label{fig:kitaev_chain_pi}
\end{figure*}

To ascertain for the physical meaning to the oscillations of the susceptibility, it is instructive to cast this quantity in  the following form:
\begin{equation}
\Pi_{tot}(\omega,\phi)=\sum_{j=0}^\infty a_{j}\cos{(2\pi j \Phi/\Phi_0+\delta\phi_j})\,,
\end{equation}
where $a_j$ are (complex) Fourier coefficients of the expansion, and $\delta\phi_j$ are phase shifts, all these quantities being functions of the chemical potential $\mu$ and the gap $\Delta$. In the topological phase, the dominant terms are $a_0$ and $a_1$, namely those corresponding to constant and $\Phi_0$ oscillations, respectively. In the topologically  trivial regime instead, $a_1=0$, and thus the usual Josephson effect dominates. Moreover, we  expect that all the coefficients with $j=2p+1$, $p=1,2,\dots$ to be zero above the topological transition (which does not host any Majoranas), while below such coefficients are in principle non-zero. We stress that the Josephson current has also a similar dependence on the applied flux: for a $p$-wave superconductor, the current $I_J(\Phi)\propto \sin\left(4\pi\Phi/\Phi_0\right)$ in the non-topological region (normal Josephson effect), while this is $I_J(\Phi)\propto \sin\left(2\pi\Phi/\Phi_0\right)$ in the topological region (fractional Josephson effect).



\subsection{Parity dependence}

In this section, we analyze the parity dependence of the electronic susceptibility, and implicitly of the cavity response. In Fig.~\ref{fig:kitaev_chain_pi}, we plot the various contributions to the susceptibility in the topological regime for the two different parities of the ground state, $f_M=0,1$. By inspecting the plots, we can conclude that there are two main features that separate the two parities: first, the oscillations (or maxima) of the susceptibilities are shifted by $\Phi_0$, and second, there is a strong asymmetry between the curves for the two parities, especially in the non-diagonal contribution. The shift by $\Phi_0$ can be explained  by means of the fractional Josephson effect, which is due to the presence of Majorana bound states that give rise to coherent  transfer of charge $e$ or $-e$, depending on the parity,  between the two bound states. That in turn gives rise to dissipationless  currents opposite in sign for the two parities, and which coincide for $\Phi=(2n+1)\Phi_0/4$, as can be seen also from the plots. The asymmetry of the Josephson contribution  can be explained by inspecting the expression for the supercurrent flow: $I_{J}=\sum_{n=bulk}i_n+(-1)^{f_M}i_M$, i.e. the bulk contribution is constant, as its parity is assumed unchanged, while the Majorana contribution depends on the parity of the ground state $f_M=0,1$. The non-diagonal susceptibility shows a much stronger asymmetry, and it is due to an interplay between the occupancy of the Majorana and the matrix elements of the current operator  $I$ between the Majorana and the bulk states, or simply by $\Pi_{BM}$ given by:
\begin{align}
\Pi_{BM}(\omega)&=-\omega\sum_{n\neq M}\frac{f_n-f_M}{\epsilon_n-\epsilon_M}\frac{|\langle n|I|M\rangle|^2}{(\epsilon_n-\epsilon_M)-\omega-i\gamma}\,,
\end{align}  
where $M$ stands for the Majorana.

\subsection{Frequency dependence}

\begin{figure}[t] 
\includegraphics[width=0.99\linewidth]{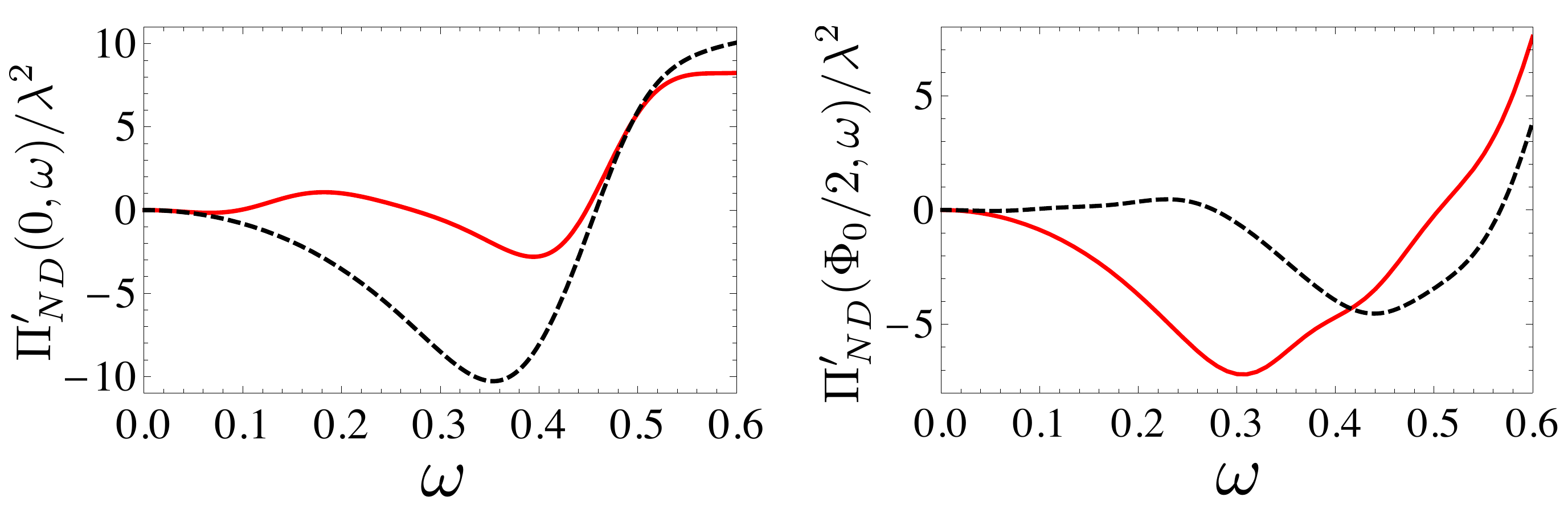}
\caption{Dependence of the real part of the  non-diagonal susceptibility on the frequency $\omega$ for two values of the external flux, $\Phi=0$ (left) and $\Phi=\Phi_0/2$ (right), and for the two parity states $f_M=0$ (dashed-black) and $f_M=1$ (full-red) lines, respectively.  The parameters for the plots are: $\mu=-0.5$,  $\Delta=0.1$, $t'=0.5$, $N = 50$, and all energies are expressed in terms of $t=1$.} 
\label{omega_dep}
\end{figure}

Finally, we discuss briefly the frequency dependence of the susceptibility (thus of the cavity response) for different parity and flux realizations, respectively. We note that the Josephson contribution is frequency independent, and the only dependence on $\omega$ arises from the non-diagonal susceptibility. In Fig.~\ref{omega_dep}, we plot the real part of the non-diagonal susceptibility as a function of the frequency in the topological regime, and for the two different parities. The non-monotonic behavior of the curves can be related to the resonance condition when the frequency $\omega$ matches the effective $p$-wave gap and the sign of the susceptibility changes. Interestingly, the dependence on $\omega$ is different for different parities $f_M=0,1$ and for different values of the magnetic flux $\Phi$. Such features can again be traced back to the dependence of $\Pi_{BM}$ on both the flux and the parity of the topological superconductor. We note, however, that the curves become identical (not shown) at flux values $\Phi=(2n+1)\Phi_0/4$, with $n=0,1,\dots$, at which instances the degeneracy of the Majorana energy levels is restored. 

\section{Conclusions}

In this paper, we studied theoretically a Kitaev ring interrupted by a weak link and pierced by a magnetic flux, and coupled inductively to a microwave cavity. We established an input-output description for the cavity transmission, and showed that the cavity response depends strongly on various electronic parameters, such as the chemical potential, the magnetic flux, and the parity of the superconducting ground state.  We found a $4\pi$ ($2\pi$) variation of the cavity response  with respect to the external magnetic flux in the topological (trivial) phase, and related such dependence to the fractional (normal) Josephson effect. As opposed to previous works, our theory takes into account, on equal footing, the low-energy Majorana modes, the gaped bulk states, and the interplay between these states in the presence of the cavity. Such a description allows one to describe not only the superconducting ring deep in the topological regime, but also the topological transition, and the crossover from the fractional to the normal Josephson effect in the presence of a magnetic flux. 

While our theory treats the Kitaev toy model, we stress that it has been shown theoretically that such a model can be emulated by the spectrum of a spin-orbit semiconducting ring subject to an external magnetic field and coupled by the proximity effect to an $s$-wave superconductor. The main difference between such a model and the Kitaev chain is the number of states, which is doubled, as one needs to account for the spin degree of freedom.  While such a study is beyond the scope of our paper, we strongly believe, based on our previous  findings in Ref.~\onlinecite{dmytruk2015cavity}, that our main results on the visualization of the transition from the fractional to the normal Josephson effect by utilizing a microwave cavity are robust. 

\begin{acknowledgements}
We would like to thank Helene Bouchiat for useful discussions. This work was supported by the French Agence Nationale de la Recherche through the contract ANR Mistral.

\end{acknowledgements}


\begin{thebibliography}{41}
\expandafter\ifx\csname natexlab\endcsname\relax\def\natexlab#1{#1}\fi
\expandafter\ifx\csname bibnamefont\endcsname\relax
  \def\bibnamefont#1{#1}\fi
\expandafter\ifx\csname bibfnamefont\endcsname\relax
  \def\bibfnamefont#1{#1}\fi
\expandafter\ifx\csname citenamefont\endcsname\relax
  \def\citenamefont#1{#1}\fi
\expandafter\ifx\csname url\endcsname\relax
  \def\url#1{\texttt{#1}}\fi
\expandafter\ifx\csname urlprefix\endcsname\relax\def\urlprefix{URL }\fi
\providecommand{\bibinfo}[2]{#2}
\providecommand{\eprint}[2][]{\url{#2}}

\bibitem[{\citenamefont{Read and Green}(2000)}]{read2000paired}
\bibinfo{author}{\bibfnamefont{N.}~\bibnamefont{Read}} \bibnamefont{and}
  \bibinfo{author}{\bibfnamefont{D.}~\bibnamefont{Green}},
  \bibinfo{journal}{Phys. Rev. B} \textbf{\bibinfo{volume}{61}},
  \bibinfo{pages}{10267} (\bibinfo{year}{2000}).

\bibitem[{\citenamefont{Ivanov}(2001)}]{ivanov2001non}
\bibinfo{author}{\bibfnamefont{D.~A.} \bibnamefont{Ivanov}},
  \bibinfo{journal}{Phys. Rev. Lett.} \textbf{\bibinfo{volume}{86}},
  \bibinfo{pages}{268} (\bibinfo{year}{2001}).

\bibitem[{\citenamefont{Kitaev}(2001)}]{kitaev2001unpaired}
\bibinfo{author}{\bibfnamefont{A.~Y.} \bibnamefont{Kitaev}},
  \bibinfo{journal}{Physics-Uspekhi} \textbf{\bibinfo{volume}{44}},
  \bibinfo{pages}{131} (\bibinfo{year}{2001}).

\bibitem{yakovenkoEPJB2004}
H.-J. Kwon, K. Sengupta, and V.M. Yakovenko,
Eur. Phys. J. B {\bf 37}, 349 (2004).

\bibitem[{\citenamefont{Fu and Kane}(2008)}]{fu2008superconducting}
\bibinfo{author}{\bibfnamefont{L.}~\bibnamefont{Fu}} \bibnamefont{and}
  \bibinfo{author}{\bibfnamefont{C.~L.} \bibnamefont{Kane}},
  \bibinfo{journal}{Phys. Rev. Lett.} \textbf{\bibinfo{volume}{100}},
  \bibinfo{pages}{096407} (\bibinfo{year}{2008}).

\bibitem[{\citenamefont{Fu and Kane}(2009)}]{fu2009josephson}
\bibinfo{author}{\bibfnamefont{L.}~\bibnamefont{Fu}} \bibnamefont{and}
  \bibinfo{author}{\bibfnamefont{C.~L.} \bibnamefont{Kane}},
  \bibinfo{journal}{Phys. Rev. B} \textbf{\bibinfo{volume}{79}},
  \bibinfo{pages}{161408} (\bibinfo{year}{2009}).

\bibitem[{\citenamefont{Hart et~al.}(2014)\citenamefont{Hart, Ren, Wagner,
  Leubner, M{\"u}hlbauer, Br{\"u}ne, Buhmann, Molenkamp, and
  Yacoby}}]{hart2014induced}
\bibinfo{author}{\bibfnamefont{S.}~\bibnamefont{Hart}},
  \bibinfo{author}{\bibfnamefont{H.}~\bibnamefont{Ren}},
  \bibinfo{author}{\bibfnamefont{T.}~\bibnamefont{Wagner}},
  \bibinfo{author}{\bibfnamefont{P.}~\bibnamefont{Leubner}},
  \bibinfo{author}{\bibfnamefont{M.}~\bibnamefont{M{\"u}hlbauer}},
  \bibinfo{author}{\bibfnamefont{C.}~\bibnamefont{Br{\"u}ne}},
  \bibinfo{author}{\bibfnamefont{H.}~\bibnamefont{Buhmann}},
  \bibinfo{author}{\bibfnamefont{L.~W.} \bibnamefont{Molenkamp}},
  \bibnamefont{and} \bibinfo{author}{\bibfnamefont{A.}~\bibnamefont{Yacoby}},
  \bibinfo{journal}{Nature Physics}  (\bibinfo{year}{2014}).

\bibitem[{\citenamefont{Oreg et~al.}(2010)\citenamefont{Oreg, Refael, and von
  Oppen}}]{oreg2010helical}
\bibinfo{author}{\bibfnamefont{Y.}~\bibnamefont{Oreg}},
  \bibinfo{author}{\bibfnamefont{G.}~\bibnamefont{Refael}}, \bibnamefont{and}
  \bibinfo{author}{\bibfnamefont{F.}~\bibnamefont{von Oppen}},
  \bibinfo{journal}{Phys. Rev. Lett.} \textbf{\bibinfo{volume}{105}},
  \bibinfo{pages}{177002} (\bibinfo{year}{2010}).

\bibitem[{\citenamefont{Lutchyn et~al.}(2010)\citenamefont{Lutchyn, Sau, and
  Sarma}}]{lutchyn2010majorana}
\bibinfo{author}{\bibfnamefont{R.~M.} \bibnamefont{Lutchyn}},
  \bibinfo{author}{\bibfnamefont{J.~D.} \bibnamefont{Sau}}, \bibnamefont{and}
  \bibinfo{author}{\bibfnamefont{S.~D.} \bibnamefont{Sarma}},
  \bibinfo{journal}{Phys. Rev. Lett.} \textbf{\bibinfo{volume}{105}},
  \bibinfo{pages}{077001} (\bibinfo{year}{2010}).

\bibitem[{\citenamefont{Mourik et~al.}(2012)\citenamefont{Mourik, Zuo, Frolov,
  Plissard, Bakkers, and Kouwenhoven}}]{mourik2012signatures}
\bibinfo{author}{\bibfnamefont{V.}~\bibnamefont{Mourik}},
  \bibinfo{author}{\bibfnamefont{K.}~\bibnamefont{Zuo}},
  \bibinfo{author}{\bibfnamefont{S.}~\bibnamefont{Frolov}},
  \bibinfo{author}{\bibfnamefont{S.}~\bibnamefont{Plissard}},
  \bibinfo{author}{\bibfnamefont{E.}~\bibnamefont{Bakkers}}, \bibnamefont{and}
  \bibinfo{author}{\bibfnamefont{L.}~\bibnamefont{Kouwenhoven}},
  \bibinfo{journal}{Science} \textbf{\bibinfo{volume}{336}},
  \bibinfo{pages}{1003} (\bibinfo{year}{2012}).

\bibitem[{\citenamefont{Albrecht et~al.}(2016)\citenamefont{Albrecht,
  Higginbotham, Madsen, Kuemmeth, Jespersen, Nyg{\aa}rd, Krogstrup, and
  Marcus}}]{albrecht2016exponential}
\bibinfo{author}{\bibfnamefont{S.}~\bibnamefont{Albrecht}},
  \bibinfo{author}{\bibfnamefont{A.}~\bibnamefont{Higginbotham}},
  \bibinfo{author}{\bibfnamefont{M.}~\bibnamefont{Madsen}},
  \bibinfo{author}{\bibfnamefont{F.}~\bibnamefont{Kuemmeth}},
  \bibinfo{author}{\bibfnamefont{T.}~\bibnamefont{Jespersen}},
  \bibinfo{author}{\bibfnamefont{J.}~\bibnamefont{Nyg{\aa}rd}},
  \bibinfo{author}{\bibfnamefont{P.}~\bibnamefont{Krogstrup}},
  \bibnamefont{and} \bibinfo{author}{\bibfnamefont{C.}~\bibnamefont{Marcus}},
  \bibinfo{journal}{Nature} \textbf{\bibinfo{volume}{531}},
  \bibinfo{pages}{206} (\bibinfo{year}{2016}).

\bibitem[{\citenamefont{Nadj-Perge et~al.}(2013)\citenamefont{Nadj-Perge,
  Drozdov, Bernevig, and Yazdani}}]{nadj2013proposal}
\bibinfo{author}{\bibfnamefont{S.}~\bibnamefont{Nadj-Perge}},
  \bibinfo{author}{\bibfnamefont{I.}~\bibnamefont{Drozdov}},
  \bibinfo{author}{\bibfnamefont{B.}~\bibnamefont{Bernevig}}, \bibnamefont{and}
  \bibinfo{author}{\bibfnamefont{A.}~\bibnamefont{Yazdani}},
  \bibinfo{journal}{Phys. Rev. B} \textbf{\bibinfo{volume}{88}},
  \bibinfo{pages}{020407} (\bibinfo{year}{2013}).

\bibitem[{\citenamefont{Klinovaja et~al.}(2013)\citenamefont{Klinovaja, Stano,
  Yazdani, and Loss}}]{klinovaja2013topological}
\bibinfo{author}{\bibfnamefont{J.}~\bibnamefont{Klinovaja}},
  \bibinfo{author}{\bibfnamefont{P.}~\bibnamefont{Stano}},
  \bibinfo{author}{\bibfnamefont{A.}~\bibnamefont{Yazdani}}, \bibnamefont{and}
  \bibinfo{author}{\bibfnamefont{D.}~\bibnamefont{Loss}},
  \bibinfo{journal}{Phys. Rev. Lett.} \textbf{\bibinfo{volume}{111}},
  \bibinfo{pages}{186805} (\bibinfo{year}{2013}).

\bibitem[{\citenamefont{Braunecker and Simon}(2013)}]{braunecker2013interplay}
\bibinfo{author}{\bibfnamefont{B.}~\bibnamefont{Braunecker}} \bibnamefont{and}
  \bibinfo{author}{\bibfnamefont{P.}~\bibnamefont{Simon}},
  \bibinfo{journal}{Phys. Rev. Lett.} \textbf{\bibinfo{volume}{111}},
  \bibinfo{pages}{147202} (\bibinfo{year}{2013}).

\bibitem[{\citenamefont{Vazifeh and Franz}(2013)}]{vazifeh2013self}
\bibinfo{author}{\bibfnamefont{M.}~\bibnamefont{Vazifeh}} \bibnamefont{and}
  \bibinfo{author}{\bibfnamefont{M.}~\bibnamefont{Franz}},
  \bibinfo{journal}{Phys. Rev. Lett.} \textbf{\bibinfo{volume}{111}},
  \bibinfo{pages}{206802} (\bibinfo{year}{2013}).

\bibitem[{\citenamefont{Pientka
  et~al.}(2013{\natexlab{a}})\citenamefont{Pientka, Glazman, and von
  Oppen}}]{pientka2013topological}
\bibinfo{author}{\bibfnamefont{F.}~\bibnamefont{Pientka}},
  \bibinfo{author}{\bibfnamefont{L.~I.} \bibnamefont{Glazman}},
  \bibnamefont{and} \bibinfo{author}{\bibfnamefont{F.}~\bibnamefont{von
  Oppen}}, \bibinfo{journal}{Phys. Rev. B} \textbf{\bibinfo{volume}{88}},
  \bibinfo{pages}{155420} (\bibinfo{year}{2013}{\natexlab{a}}).

\bibitem[{\citenamefont{Kim et~al.}(2014)\citenamefont{Kim, Cheng, Bauer,
  Lutchyn, and Sarma}}]{kim2014helical}
\bibinfo{author}{\bibfnamefont{Y.}~\bibnamefont{Kim}},
  \bibinfo{author}{\bibfnamefont{M.}~\bibnamefont{Cheng}},
  \bibinfo{author}{\bibfnamefont{B.}~\bibnamefont{Bauer}},
  \bibinfo{author}{\bibfnamefont{R.~M.} \bibnamefont{Lutchyn}},
  \bibnamefont{and} \bibinfo{author}{\bibfnamefont{S.~D.} \bibnamefont{Sarma}},
  \bibinfo{journal}{Phys. Rev. B} \textbf{\bibinfo{volume}{90}},
  \bibinfo{pages}{060401} (\bibinfo{year}{2014}).

\bibitem[{\citenamefont{Nadj-Perge et~al.}(2014)\citenamefont{Nadj-Perge,
  Drozdov, Li, Chen, Jeon, Seo, MacDonald, Bernevig, and
  Yazdani}}]{nadj2014observation}
\bibinfo{author}{\bibfnamefont{S.}~\bibnamefont{Nadj-Perge}},
  \bibinfo{author}{\bibfnamefont{I.~K.} \bibnamefont{Drozdov}},
  \bibinfo{author}{\bibfnamefont{J.}~\bibnamefont{Li}},
  \bibinfo{author}{\bibfnamefont{H.}~\bibnamefont{Chen}},
  \bibinfo{author}{\bibfnamefont{S.}~\bibnamefont{Jeon}},
  \bibinfo{author}{\bibfnamefont{J.}~\bibnamefont{Seo}},
  \bibinfo{author}{\bibfnamefont{A.~H.} \bibnamefont{MacDonald}},
  \bibinfo{author}{\bibfnamefont{B.~A.} \bibnamefont{Bernevig}},
  \bibnamefont{and} \bibinfo{author}{\bibfnamefont{A.}~\bibnamefont{Yazdani}},
  \bibinfo{journal}{Science} \textbf{\bibinfo{volume}{346}},
  \bibinfo{pages}{602} (\bibinfo{year}{2014}).


\bibitem[{\citenamefont{Alicea}(2012)}]{alicea2012new}
\bibinfo{author}{\bibfnamefont{J.}~\bibnamefont{Alicea}},
  \bibinfo{journal}{Reports on Progress in Physics}
  \textbf{\bibinfo{volume}{75}}, \bibinfo{pages}{076501}
  (\bibinfo{year}{2012}).

\bibitem[{\citenamefont{Pientka
  et~al.}(2013{\natexlab{b}})\citenamefont{Pientka, Romito, Duckheim, Oreg, and
  von Oppen}}]{pientka2013signatures}
\bibinfo{author}{\bibfnamefont{F.}~\bibnamefont{Pientka}},
  \bibinfo{author}{\bibfnamefont{A.}~\bibnamefont{Romito}},
  \bibinfo{author}{\bibfnamefont{M.}~\bibnamefont{Duckheim}},
  \bibinfo{author}{\bibfnamefont{Y.}~\bibnamefont{Oreg}}, \bibnamefont{and}
  \bibinfo{author}{\bibfnamefont{F.}~\bibnamefont{von Oppen}},
  \bibinfo{journal}{New Journal of Physics} \textbf{\bibinfo{volume}{15}},
  \bibinfo{pages}{025001} (\bibinfo{year}{2013}{\natexlab{b}}).

\bibitem[{\citenamefont{Rokhinson et~al.}(2012)\citenamefont{Rokhinson, Liu,
  and Furdyna}}]{rokhinson2012fractional}
\bibinfo{author}{\bibfnamefont{L.~P.} \bibnamefont{Rokhinson}},
  \bibinfo{author}{\bibfnamefont{X.}~\bibnamefont{Liu}}, \bibnamefont{and}
  \bibinfo{author}{\bibfnamefont{J.~K.} \bibnamefont{Furdyna}},
  \bibinfo{journal}{Nature Physics} \textbf{\bibinfo{volume}{8}},
  \bibinfo{pages}{795} (\bibinfo{year}{2012}).

\bibitem[{\citenamefont{Wiedenmann et~al.}(2016)\citenamefont{Wiedenmann,
  Bocquillon, Deacon, Hartinger, Herrmann, Klapwijk, Maier, Ames, Br{\"u}ne,
  Gould et~al.}}]{wiedenmann2016}
\bibinfo{author}{\bibfnamefont{J.}~\bibnamefont{Wiedenmann}},
  \bibinfo{author}{\bibfnamefont{E.}~\bibnamefont{Bocquillon}},
  \bibinfo{author}{\bibfnamefont{R.~S.} \bibnamefont{Deacon}},
  \bibinfo{author}{\bibfnamefont{S.}~\bibnamefont{Hartinger}},
  \bibinfo{author}{\bibfnamefont{O.}~\bibnamefont{Herrmann}},
  \bibinfo{author}{\bibfnamefont{T.~M.} \bibnamefont{Klapwijk}},
  \bibinfo{author}{\bibfnamefont{L.}~\bibnamefont{Maier}},
  \bibinfo{author}{\bibfnamefont{C.}~\bibnamefont{Ames}},
  \bibinfo{author}{\bibfnamefont{C.}~\bibnamefont{Br{\"u}ne}},
  \bibinfo{author}{\bibfnamefont{C.}~\bibnamefont{Gould}},
  \bibnamefont{et~al.}, \bibinfo{journal}{Nature Communications}
  \textbf{\bibinfo{volume}{7}} (\bibinfo{year}{2016}).

\bibitem[{\citenamefont{Della~Rocca et~al.}(2007)\citenamefont{Della~Rocca,
  Chauvin, Huard, Pothier, Esteve, and Urbina}}]{della2007measurement}
\bibinfo{author}{\bibfnamefont{M.}~\bibnamefont{Della~Rocca}},
  \bibinfo{author}{\bibfnamefont{M.}~\bibnamefont{Chauvin}},
  \bibinfo{author}{\bibfnamefont{B.}~\bibnamefont{Huard}},
  \bibinfo{author}{\bibfnamefont{H.}~\bibnamefont{Pothier}},
  \bibinfo{author}{\bibfnamefont{D.}~\bibnamefont{Esteve}}, \bibnamefont{and}
  \bibinfo{author}{\bibfnamefont{C.}~\bibnamefont{Urbina}},
  \bibinfo{journal}{Phys. Rev. Lett.} \textbf{\bibinfo{volume}{99}},
  \bibinfo{pages}{127005} (\bibinfo{year}{2007}).

\bibitem[{\citenamefont{Trif and Tserkovnyak}(2012)}]{trif2012resonantly}
\bibinfo{author}{\bibfnamefont{M.}~\bibnamefont{Trif}} \bibnamefont{and}
  \bibinfo{author}{\bibfnamefont{Y.}~\bibnamefont{Tserkovnyak}},
  \bibinfo{journal}{Phys. Rev. Lett.} \textbf{\bibinfo{volume}{109}},
  \bibinfo{pages}{257002} (\bibinfo{year}{2012}).

\bibitem[{\citenamefont{Schmidt
  et~al.}(2013{\natexlab{a}})\citenamefont{Schmidt, Nunnenkamp, and
  Bruder}}]{schmidt2013majorana}
\bibinfo{author}{\bibfnamefont{T.~L.} \bibnamefont{Schmidt}},
  \bibinfo{author}{\bibfnamefont{A.}~\bibnamefont{Nunnenkamp}},
  \bibnamefont{and} \bibinfo{author}{\bibfnamefont{C.}~\bibnamefont{Bruder}},
  \bibinfo{journal}{Phys. Rev. Lett.} \textbf{\bibinfo{volume}{110}},
  \bibinfo{pages}{107006} (\bibinfo{year}{2013}{\natexlab{a}}).

\bibitem[{\citenamefont{Schmidt
  et~al.}(2013{\natexlab{b}})\citenamefont{Schmidt, Nunnenkamp, and
  Bruder}}]{schmidt2013njp}
\bibinfo{author}{\bibfnamefont{T.~L.} \bibnamefont{Schmidt}},
  \bibinfo{author}{\bibfnamefont{A.}~\bibnamefont{Nunnenkamp}},
  \bibnamefont{and} \bibinfo{author}{\bibfnamefont{C.}~\bibnamefont{Bruder}},
  \bibinfo{journal}{New Journal of Physics} \textbf{\bibinfo{volume}{15}},
  \bibinfo{pages}{025043} (\bibinfo{year}{2013}{\natexlab{b}}).

\bibitem[{\citenamefont{Cottet et~al.}(2013)\citenamefont{Cottet, Kontos, and
  Dou{\c{c}}ot}}]{cottet2013squeezing}
\bibinfo{author}{\bibfnamefont{A.}~\bibnamefont{Cottet}},
  \bibinfo{author}{\bibfnamefont{T.}~\bibnamefont{Kontos}}, \bibnamefont{and}
  \bibinfo{author}{\bibfnamefont{B.}~\bibnamefont{Dou{\c{c}}ot}},
  \bibinfo{journal}{Phys. Rev. B} \textbf{\bibinfo{volume}{88}},
  \bibinfo{pages}{195415} (\bibinfo{year}{2013}).

\bibitem[{\citenamefont{M{\"u}ller et~al.}(2013)\citenamefont{M{\"u}ller,
  Bourassa, and Blais}}]{muller2013detection}
\bibinfo{author}{\bibfnamefont{C.}~\bibnamefont{M{\"u}ller}},
  \bibinfo{author}{\bibfnamefont{J.}~\bibnamefont{Bourassa}}, \bibnamefont{and}
  \bibinfo{author}{\bibfnamefont{A.}~\bibnamefont{Blais}},
  \bibinfo{journal}{Phys. Rev. B} \textbf{\bibinfo{volume}{88}},
  \bibinfo{pages}{235401} (\bibinfo{year}{2013}).

\bibitem{XuePRA2013}
Zheng-Yuan Xue, L. B. Shao, Yong Hu, Shi-Liang Zhu, and Z. D. Wang,
Phys. Rev. A {\bf 88}, 024303 (2013).

\bibitem[{\citenamefont{Ohm and Hassler}(2014)}]{ohm2014majorana}
\bibinfo{author}{\bibfnamefont{C.}~\bibnamefont{Ohm}} \bibnamefont{and}
  \bibinfo{author}{\bibfnamefont{F.}~\bibnamefont{Hassler}},
  \bibinfo{journal}{New Journal of Physics} \textbf{\bibinfo{volume}{16}},
  \bibinfo{pages}{015009} (\bibinfo{year}{2014}).

\bibitem{XueSciRep2015}
Zheng-Yuan Xue, Ming Gong, Jia Li, Yong Hu, Shi-Liang Zhu, and Z. D. Wang
Sci. Rep. {\bf 5}  12233; doi: 10.1038/srep12233 (2015).

\bibitem[{\citenamefont{Yavilberg et~al.}(2015)\citenamefont{Yavilberg,
  Ginossar, and Grosfeld}}]{yavilberg2015fermion}
\bibinfo{author}{\bibfnamefont{K.}~\bibnamefont{Yavilberg}},
  \bibinfo{author}{\bibfnamefont{E.}~\bibnamefont{Ginossar}}, \bibnamefont{and}
  \bibinfo{author}{\bibfnamefont{E.}~\bibnamefont{Grosfeld}},
  \bibinfo{journal}{Phys. Rev. B} \textbf{\bibinfo{volume}{92}},
  \bibinfo{pages}{075143} (\bibinfo{year}{2015}).

\bibitem[{\citenamefont{Ohm and Hassler}(2015)}]{ohm2015microwave}
\bibinfo{author}{\bibfnamefont{C.}~\bibnamefont{Ohm}} \bibnamefont{and}
  \bibinfo{author}{\bibfnamefont{F.}~\bibnamefont{Hassler}},
  \bibinfo{journal}{Phys. Rev. B} \textbf{\bibinfo{volume}{91}},
  \bibinfo{pages}{085406} (\bibinfo{year}{2015}).

\bibitem[{\citenamefont{Dmytruk et~al.}(2015)\citenamefont{Dmytruk, Trif, and
  Simon}}]{dmytruk2015cavity}
\bibinfo{author}{\bibfnamefont{O.}~\bibnamefont{Dmytruk}},
  \bibinfo{author}{\bibfnamefont{M.}~\bibnamefont{Trif}}, \bibnamefont{and}
  \bibinfo{author}{\bibfnamefont{P.}~\bibnamefont{Simon}},
  \bibinfo{journal}{Phys. Rev. B} \textbf{\bibinfo{volume}{92}},
  \bibinfo{pages}{245432} (\bibinfo{year}{2015}).

\bibitem[{\citenamefont{Blais et~al.}(2004)\citenamefont{Blais, Huang,
  Wallraff, Girvin, and Schoelkopf}}]{blais2004cavity}
\bibinfo{author}{\bibfnamefont{A.}~\bibnamefont{Blais}},
  \bibinfo{author}{\bibfnamefont{R.-S.} \bibnamefont{Huang}},
  \bibinfo{author}{\bibfnamefont{A.}~\bibnamefont{Wallraff}},
  \bibinfo{author}{\bibfnamefont{S.}~\bibnamefont{Girvin}}, \bibnamefont{and}
  \bibinfo{author}{\bibfnamefont{R.~J.} \bibnamefont{Schoelkopf}},
  \bibinfo{journal}{Phys. Rev. A} \textbf{\bibinfo{volume}{69}},
  \bibinfo{pages}{062320} (\bibinfo{year}{2004}).

\bibitem[{\citenamefont{Wallraff et~al.}(2004)\citenamefont{Wallraff, Schuster,
  Blais, Frunzio, Huang, Majer, Kumar, Girvin, and
  Schoelkopf}}]{wallraff2004strong}
\bibinfo{author}{\bibfnamefont{A.}~\bibnamefont{Wallraff}},
  \bibinfo{author}{\bibfnamefont{D.~I.} \bibnamefont{Schuster}},
  \bibinfo{author}{\bibfnamefont{A.}~\bibnamefont{Blais}},
  \bibinfo{author}{\bibfnamefont{L.}~\bibnamefont{Frunzio}},
  \bibinfo{author}{\bibfnamefont{R.-S.} \bibnamefont{Huang}},
  \bibinfo{author}{\bibfnamefont{J.}~\bibnamefont{Majer}},
  \bibinfo{author}{\bibfnamefont{S.}~\bibnamefont{Kumar}},
  \bibinfo{author}{\bibfnamefont{S.~M.} \bibnamefont{Girvin}},
  \bibnamefont{and} \bibinfo{author}{\bibfnamefont{R.~J.}
  \bibnamefont{Schoelkopf}}, \bibinfo{journal}{Nature}
  \textbf{\bibinfo{volume}{431}}, \bibinfo{pages}{162} (\bibinfo{year}{2004}).

\bibitem[{\citenamefont{Samkharadze et~al.}(2015)\citenamefont{Samkharadze,
  Bruno, Scarlino, Zheng, DiVincenzo, DiCarlo, and
  Vandersypen}}]{samkharadze2015high}
\bibinfo{author}{\bibfnamefont{N.}~\bibnamefont{Samkharadze}},
  \bibinfo{author}{\bibfnamefont{A.}~\bibnamefont{Bruno}},
  \bibinfo{author}{\bibfnamefont{P.}~\bibnamefont{Scarlino}},
  \bibinfo{author}{\bibfnamefont{G.}~\bibnamefont{Zheng}},
  \bibinfo{author}{\bibfnamefont{D.}~\bibnamefont{DiVincenzo}},
  \bibinfo{author}{\bibfnamefont{L.}~\bibnamefont{DiCarlo}}, \bibnamefont{and}
  \bibinfo{author}{\bibfnamefont{L.}~\bibnamefont{Vandersypen}},
  \bibinfo{journal}{arXiv preprint arXiv:1511.01760}  (\bibinfo{year}{2015}).

\bibitem[{\citenamefont{Mendes and Mora}(2015)}]{mendes2015cavity}
\bibinfo{author}{\bibfnamefont{U.~C.} \bibnamefont{Mendes}} \bibnamefont{and}
  \bibinfo{author}{\bibfnamefont{C.}~\bibnamefont{Mora}}, \bibinfo{journal}{New
  Journal of Physics} \textbf{\bibinfo{volume}{17}}, \bibinfo{pages}{113014}
  (\bibinfo{year}{2015}).

\bibitem[{\citenamefont{Dmytruk et~al.}(2016)\citenamefont{Dmytruk, Trif, Mora,
  and Simon}}]{dmytruk2016out}
\bibinfo{author}{\bibfnamefont{O.}~\bibnamefont{Dmytruk}},
  \bibinfo{author}{\bibfnamefont{M.}~\bibnamefont{Trif}},
  \bibinfo{author}{\bibfnamefont{C.}~\bibnamefont{Mora}}, \bibnamefont{and}
  \bibinfo{author}{\bibfnamefont{P.}~\bibnamefont{Simon}},
  \bibinfo{journal}{Phys. Rev. B} \textbf{\bibinfo{volume}{93}},
  \bibinfo{pages}{075425} (\bibinfo{year}{2016}).

\bibitem[{\citenamefont{Clerk et~al.}(2010)\citenamefont{Clerk, Devoret,
  Girvin, Marquardt, and Schoelkopf}}]{clerk2010introduction}
\bibinfo{author}{\bibfnamefont{A.}~\bibnamefont{Clerk}},
  \bibinfo{author}{\bibfnamefont{M.}~\bibnamefont{Devoret}},
  \bibinfo{author}{\bibfnamefont{S.}~\bibnamefont{Girvin}},
  \bibinfo{author}{\bibfnamefont{F.}~\bibnamefont{Marquardt}},
  \bibnamefont{and}
  \bibinfo{author}{\bibfnamefont{R.}~\bibnamefont{Schoelkopf}},
  \bibinfo{journal}{Reviews of Modern Physics} \textbf{\bibinfo{volume}{82}},
  \bibinfo{pages}{1155} (\bibinfo{year}{2010}).

\bibitem[{\citenamefont{Trivedi and Browne}(1988)}]{trivedi1988mesoscopic}
\bibinfo{author}{\bibfnamefont{N.}~\bibnamefont{Trivedi}} \bibnamefont{and}
  \bibinfo{author}{\bibfnamefont{D.~A.} \bibnamefont{Browne}},
  \bibinfo{journal}{Phys. Rev. B} \textbf{\bibinfo{volume}{38}},
  \bibinfo{pages}{9581} (\bibinfo{year}{1988}).

\bibitem[{\citenamefont{Ferrier et~al.}(2013)\citenamefont{Ferrier,
  Dassonneville, Gu{\'e}ron, and Bouchiat}}]{ferrier2013phase}
\bibinfo{author}{\bibfnamefont{M.}~\bibnamefont{Ferrier}},
  \bibinfo{author}{\bibfnamefont{B.}~\bibnamefont{Dassonneville}},
  \bibinfo{author}{\bibfnamefont{S.}~\bibnamefont{Gu{\'e}ron}},
  \bibnamefont{and} \bibinfo{author}{\bibfnamefont{H.}~\bibnamefont{Bouchiat}},
  \bibinfo{journal}{Phys. Rev. B} \textbf{\bibinfo{volume}{88}},
  \bibinfo{pages}{174505} (\bibinfo{year}{2013}).

\bibitem[{\citenamefont{Dassonneville et~al.}(2013)\citenamefont{Dassonneville,
  Ferrier, Gu{\'e}ron, and Bouchiat}}]{dassonneville2013dissipation}
\bibinfo{author}{\bibfnamefont{B.}~\bibnamefont{Dassonneville}},
  \bibinfo{author}{\bibfnamefont{M.}~\bibnamefont{Ferrier}},
  \bibinfo{author}{\bibfnamefont{S.}~\bibnamefont{Gu{\'e}ron}},
  \bibnamefont{and} \bibinfo{author}{\bibfnamefont{H.}~\bibnamefont{Bouchiat}},
  \bibinfo{journal}{Phys. Rev. Lett.} \textbf{\bibinfo{volume}{110}},
  \bibinfo{pages}{217001} (\bibinfo{year}{2013}).

\bibitem[{\citenamefont{Sticlet and Cayssol}(2014)}]{sticlet2014dynamical}
\bibinfo{author}{\bibfnamefont{D.}~\bibnamefont{Sticlet}} \bibnamefont{and}
  \bibinfo{author}{\bibfnamefont{J.}~\bibnamefont{Cayssol}},
  \bibinfo{journal}{Phys. Rev. B} \textbf{\bibinfo{volume}{90}},
  \bibinfo{pages}{201303} (\bibinfo{year}{2014}).

\end{thebibliography}

\end{document}